\documentclass[prl,twocolumn]{revtex4}
\usepackage{enumerate}
\usepackage{amsfonts,amssymb,amsmath}
\usepackage[]{graphics,graphicx,epsfig}
\usepackage{amsthm}
\usepackage{float}

\usepackage{pdfpages}
\usepackage{epstopdf}
\DeclareGraphicsExtensions{.png,.pdf}
\usepackage{graphicx}
\usepackage{dcolumn}
\usepackage{natbib}
\usepackage{color}
\usepackage{multirow}
\usepackage{ulem}
\newcommand{\ket}[1]{|{#1}\rangle}
\newcommand{\bra}[1]{\langle{#1}|}

\usepackage{array}
\newcolumntype{P}[1]{>{\centering\arraybackslash}p{#1}}

\begin{document}

\title{Revisiting quantum walk advantages: A mean hitting time perspective}

\author{Jan W{\'o}jcik}
\affiliation{Institute of Theoretical Physics and Astrophysics, University of Gdańsk, 80-308 Gda\'nsk, Poland}
\date{\today}

\begin{abstract}

The mean squared displacement has been widely used as the primary metric for comparing quantum and classical random walks, with quantum walks showing quadratic scaling versus linear scaling for classical walks. However, this comparison may not capture the full picture: while the mean squared displacement is well-suited for Gaussian distributions, quantum walk distributions exhibit distinctly non-Gaussian features. We propose that the mean hitting time offers a complementary perspective with clear operational meaning for search algorithms. Through analytical calculations, we show that quantum and classical walks yield identical MHT for symmetric initial conditions with two detectors, suggesting that the apparent quantum advantage seen in MSD comparisons may be context-dependent. Interestingly, introducing stochastic resetting reveals new dynamics. We demonstrate analytically that quantum walks can achieve reduced MHT under stochastic reset through quasi-momentum redistribution, while classical walks see no benefit. This quantum advantage naturally degrades with noise, the quantum walk converges to classical behavior. We suggest that MHT reduction under stochastic reset can serve as an additional signature of quantum behavior, particularly useful for characterizing quantum walk implementations on noisy quantum devices. Our results indicate that different metrics can reveal different aspects of quantum-classical comparisons in walk-based algorithms.

\end{abstract}

\maketitle
\section{Introduction}

The discrete time random walk (DTRW), first introduced by Pearson and Rayleigh \cite{rayleigh1905problem,pearson1905problem_1,pearson1905problem_2}, describes the evolution of a particle making consecutive random jumps on a lattice. This at first glance simple model quickly climbed the popularity ladder to become one of the most widely studied stochastic systems. Initially applied to describe diffusive processes in physics, it was rapidly adapted to other fields \cite{chandrasekhar1943stochastic, hughes1995random1}. One of the most every day use adaptations occurred in computer science \cite{8911513}, particularly in algorithms such as PageRank \cite{ilprints422}.

The DTRW is most commonly characterized by two measures. Since the process models diffusion, the particle's probability distribution tends toward a Gaussian, and as such can be intuitively parametrized by the mean squared displacement (MSD), which scales linearly with time $\text{MSD}\propto t$. This measure is intuitive because of the Gaussian nature of the distribution and gives a clear answer to the question of how fast would the ensemble of particle spread. However, one can look for a different measure of this process, one that captures the search-algorithmic essence of the model: namely, the mean hitting time (MHT). The MHT is the average time at which a particle will be measured at a given destination, which also intuitively provides a measure of the particle's "speed".  This simple measure can lead to important yet unintuitive results, such as the divergence of MHT on a line. Consider a random walk on an infinite chain starting from position $x_0$: the MHT to a target position $x_t$ is infinite, regardless of the target position.
\begin{figure}[t]
\includegraphics[width=8cm]{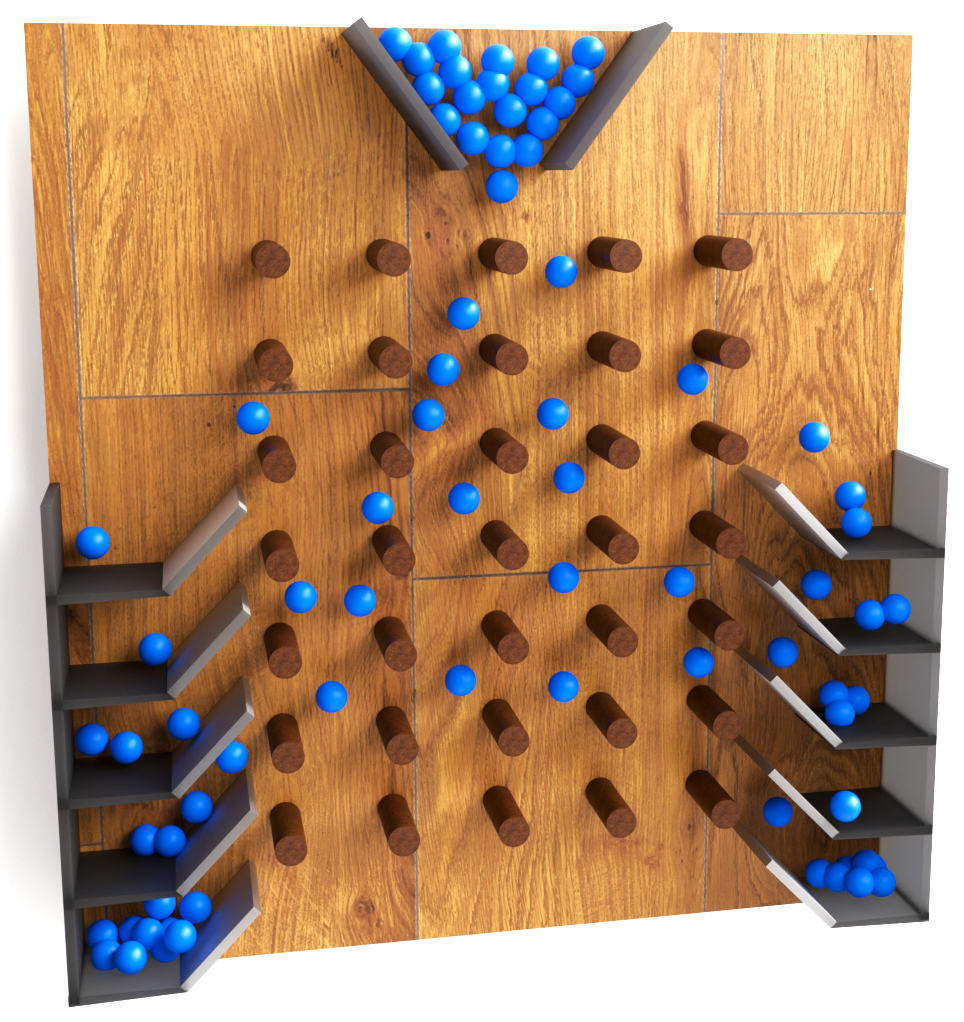}
\caption{ The modified Galton board experiment scheme. The particles (balls) are falling from the top and bouncing of the wooden pins either to the left or to the right. At the ends of the board we have shelves which trap the particles acting as detectors. They measure at what time the particle went to the target position. Repeating this experiment with many particle will give us the average time of the particle being measured at target positions so the MHT.} 
\label{galton}
\end{figure}
This divergence can be overcome in several ways. First, if one chooses two target positions (effectively making the model finite), the MHT converges. We propose a simple experimental realization of the MHT measurement using a modified Galton board, as depicted in Fig. \ref{galton}. In contrast to the standard Galton board which uses reflective walls at the boundaries, our modification replaces these walls with collection shelves that act as detectors. These detectors record the arrival time of particles reaching either end of the board. By averaging over many particle trajectories, we can directly measure the MHT experimentally. A second approach uses a stochastic reset procedure, which restarts the process to its initial condition. Using such a scheme, we once again obtain a finite MHT. For the purpose of this work, we investigate models with two target positions and explore how a stochastic reset mechanism can optimize the already finite MHT.

With the growing number of applications of classical random walks, a natural question emerged: is quantization of this model possible, potentially exploiting quantum advantage in the realm of search algorithms? Aharonov, Davidovich, and Zugary \cite{Aharonov1993} were the first to answer this question. They developed a model named the quantum walk, which soon became a toy model for many interesting and complex physical phenomena \cite{Debbasch2012,Genske2013,Cedzich2013,Cedzich2019,Chandrashekar2008,Crespi2013,Jolly2023,Wojcik2004,Zhang2016,Harper2020,Kitagawa2012,Kitagawa2010,Asboth2012}, as well as a cornerstone of numerous quantum algorithms \cite{Farhi1998,Kadian2021,Montanaro2016,Portugal2013}. It is worth mentioning that this model has also been studied experimentally across various platforms \cite{Schmitz2009,Zahringer:2010te,Flurin:2017vp,Yan:2019ve,Du2003,Ryan2005,Karski2009,Dadras2018,Do2005,Cardano:2016aa,Perets2008,Tang2018}, including quantum computers \cite{Pan2021}.

Naturally, because of its origin, the quantum walk is commonly compared to its predecessor, the classical random walk. This comparison is usually done via the MSD. One can show that the MSD of a quantum walk starting from a localized position scales quadratically with time. Based on this, many have postulated a speed-up of quantum walks over classical walks.

In this work, we explore whether MSD tells the complete story when comparing the speed of quantum and classical walks, and we investigate how MHT can provide additional insights into this comparison. Furthermore, we build upon recent findings \cite{Chelminiak2025}, providing analytical proofs of quantum advantage in models with stochastic reset. Finally, we suggest possible applications of this method to algorithms in noisy environments.

\begin{figure}[t]
\includegraphics[width=8cm]{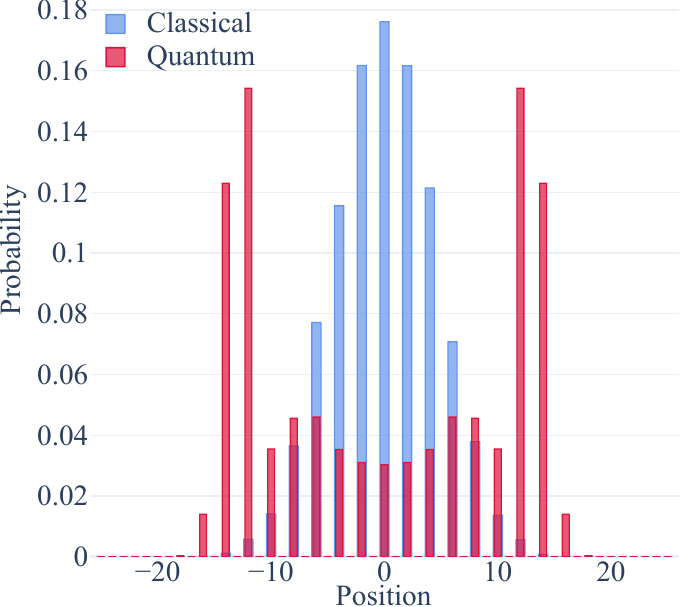}
\caption{ The spatial probability distributions of the classical (blue) and quantum (red) walks on chain.}
\label{pdf}
\end{figure}
\section{Model}
In this paper we focus on discrete time models on chains. We start with the classical random walk which simulates dynamics of the particle. Each step of the evolution the particle will with equal probability go to one of the two neighboring positions.
In such dynamics the particle performs a diffusive motion resulting in Gaussian distribution. Thus the most used characterization of the state of the particle is the MSD, which for initially localized particle scales linearly with time 
\begin{equation}
    \text{MSD}(t) \propto t.
\end{equation}
Since the probability distribution of such walk is Gaussian, such characterization has great meaning. However, equally important from operational point of view is the MHT. To obtain it let us choose two target positions (effectively making the chain finite) and ask for the average time in which the particle will be measured by the devices (see Fig. \ref{galton}). The MHT then is given by 
\begin{equation}
    \text{MHT} = \sum_{t=0}^\infty tp(t),
\end{equation}

This problem was extensively studied in context of the famous Ruin Game by Feller in \cite{Feller1968}. For particle initially localized at vertex $x_0$ the MHT simplifies to the multiplication of distances from the initial position to target positions
\begin{equation}
    \text{MHT} (x_0,x_{t_1},x_{t_2}) = |(x_{t_1}-x_0)(x_{t_2}-x_0)|,
\end{equation}
with initial position $x_0$ and target positions $x_{t_i}$. Note that if one chooses the symmetric initial position, the resulting MHT is just equal to $\frac{L^2}{4}$ with $L$ being the effective length of the chain. 

Having in mind great number of applications of the classical walks to the algorithmic tasks we move to its quantum counterpart with the intent of comparing the two models. In the discrete time quantum walk on a chain we describe the state of the particle by state vector $\ket{\Psi}\in \mathcal{H}$. Additionally the particle has now not only the spatial degree of but also the two dimensional internal "coin" degree of freedom . Thus the state of the particle can be expressed as
\begin{equation}
    \ket{\Psi(t)} = \sum_x \ket{x} \otimes (a_x(t) \ket{+1} + b_x(t) \ket{-1}).
\end{equation}
The evolution is discrete in time
\begin{equation}
    \ket{\Psi(t+1)} = U\ket{\Psi(t)} ,
\end{equation}
governed by unitary evolution operator
\begin{equation}
    U =SC,
\end{equation}
with $S$ being the "step" operator that performs "coin" dependent jump
\begin{equation}
    S\ket{x,\pm1} = \ket{x\pm1,\pm1},
\end{equation}
and $C$ being the "coin" operator that performs unitary transformation of the internal degree of freedom
\begin{equation}
    C = I_x \otimes \frac{1}{\sqrt{2}}\begin{pmatrix}
        1&-1\\1&1
    \end{pmatrix}.
\end{equation}
Note that this transformation of the internal degree of freedom can be performed by any $\mathcal{U}(2)$ matrix. However, for purpose of this work we choose one of the most popular operators, namely the rotation $e^{-i\frac{\pi}{4}\sigma_y}$ of the internal degree of freedom on the Bloch sphere. 
\begin{figure}[t]
\includegraphics[width=7cm]{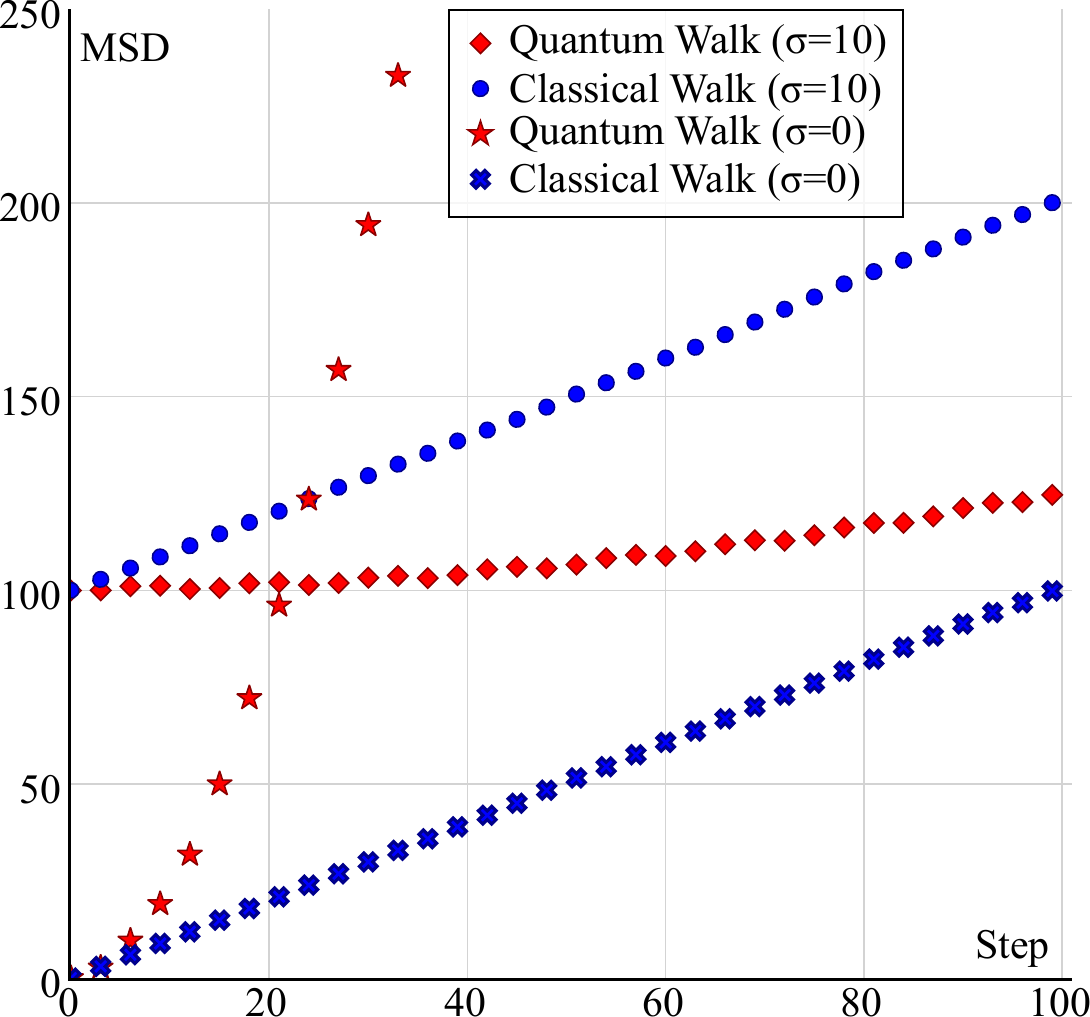}
\caption{ The MSD over time for classical (blue) and quantum (red) walk starting from initial gaussian distributions with widths $\sigma = 10$ (dots and diamonds) and $\sigma =0$ (stars and crosses). Additionally the quantum walk has been initialized with "symmetric" coin state i.e. $\frac{1}{\sqrt{2}}(\ket{+1}+i\ket{-1})$.}
\label{msd1}
\end{figure}
The probability distribution of a quantum walk after 20 steps starting from localized state $\ket{\Psi(t=0)} = \ket{x=0}\otimes \frac{1}{\sqrt{2}}(\ket{+1} + i\ket{-1})$ is depicted on Fig. \ref{pdf}. One can also calculate the MSD for such dynamics and find that it scales quadratically with time. 

The MSD quickly became the standard metric for comparing quantum and classical walks. To be specific MSD for initially localized states. Based on that result the quantum walks were claimed "faster" then classical walks \cite{Kempe:2003aa}. Specifically hinting that since classical random walks are used in overwhelming number of search algorithms then there is a good chance that quantum walks will perform even better on quantum devices. 

\begin{figure}[b]
\includegraphics[width=7cm]{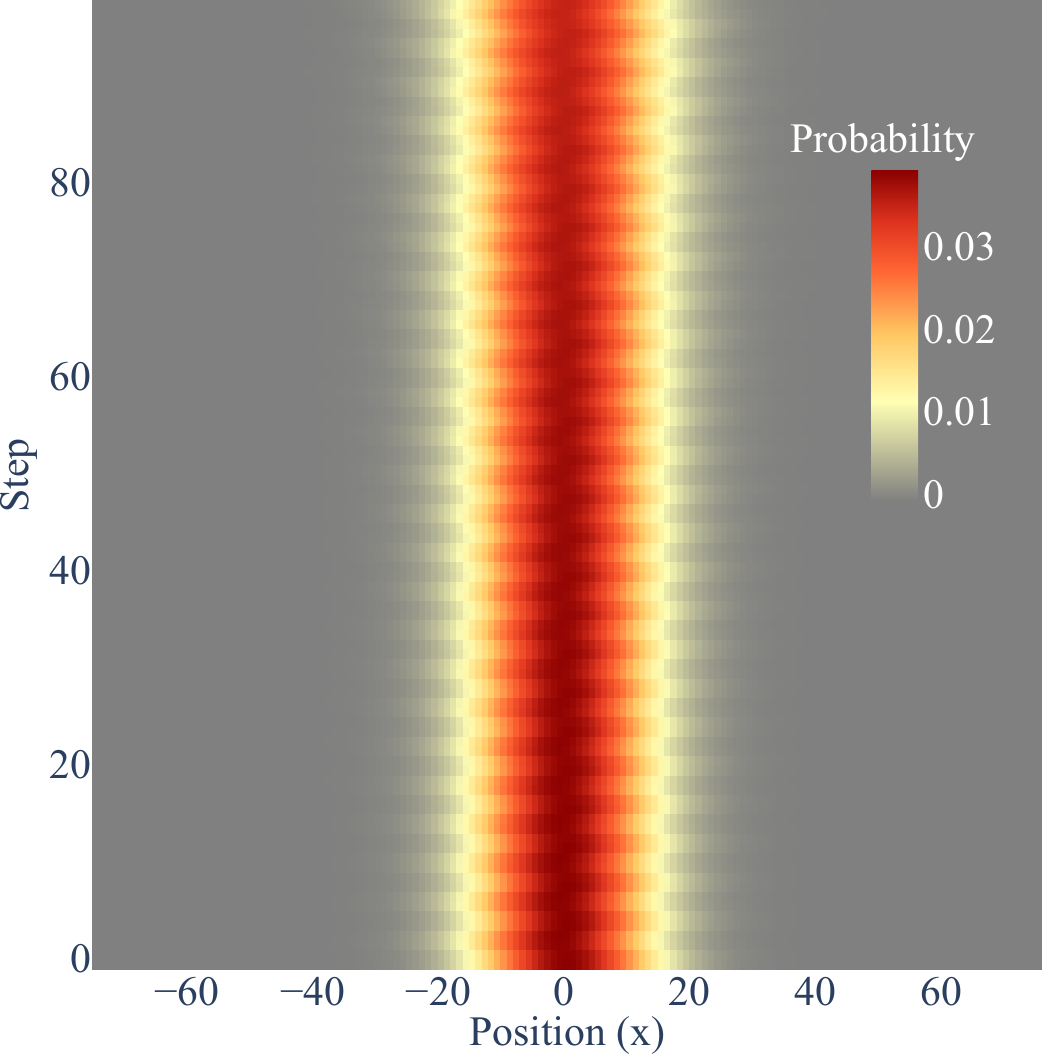}
\caption{ The probability distribution of particle during quantum walk evolution starting from initial state $\ket{\Psi} = N\sum_x e^{-(\frac{x}{2\sigma})^2}\ket{x,+1}$.} 
\label{msd2}
\end{figure}

This comparison may not tell the full story for two reasons: first, MSD is particularly well-suited for Gaussian probability distributions, while quantum walk distributions have distinctive non-Gaussian features (see Fig. \ref{pdf}), second, the MSD is indeed greater for quantum walks but for specific states. On Fig. \ref{msd1} we show how the MSD depends on time for classical and quantum walks starting from initial gaussian distributions with different widths for time interval of hundred steps. One clearly sees that indeed quantum walk has greater MSD if started from localized state ($\sigma=0$). However, once we initialize from wider distribution ($\sigma=10$ as an example) suddenly the classical walks have greater MSD (see \cite{orthey2017spreadingquantumwalksstarting}). For this particular gaussian like states the long time behavior is still ballistic but in the chosen time interval classical walks win over quantum walks. Moreover on Fig. \ref{msd2} we present how the evolution from wider distribution looks like in quantum walk. One clearly see that the spread is extremely slow. Thus we believe there is value in exploring complementary comparison methods.

We propose to examine an additional method (based on MHT) that offers a different perspective and has clear operational meaning. It is clear that the average time the particle will take until reaching the target destination is an intuitive measure of the speed of the model (specifically meaningful in context of search algorithms). 

In classical random walk the MHT is well known measure. However, for quantum walks it is quiet overlooked. It was first used by Krovi and Brun \cite{PhysRevA.74.042334}, and recently in \cite{BACH2004562,PhysRevA.78.022324,PhysRevA.73.032341,Chelminiak2025}. We will build upon their findings to compare classical and quantum walks. 

\begin{figure}[t]
\includegraphics[width=8cm]{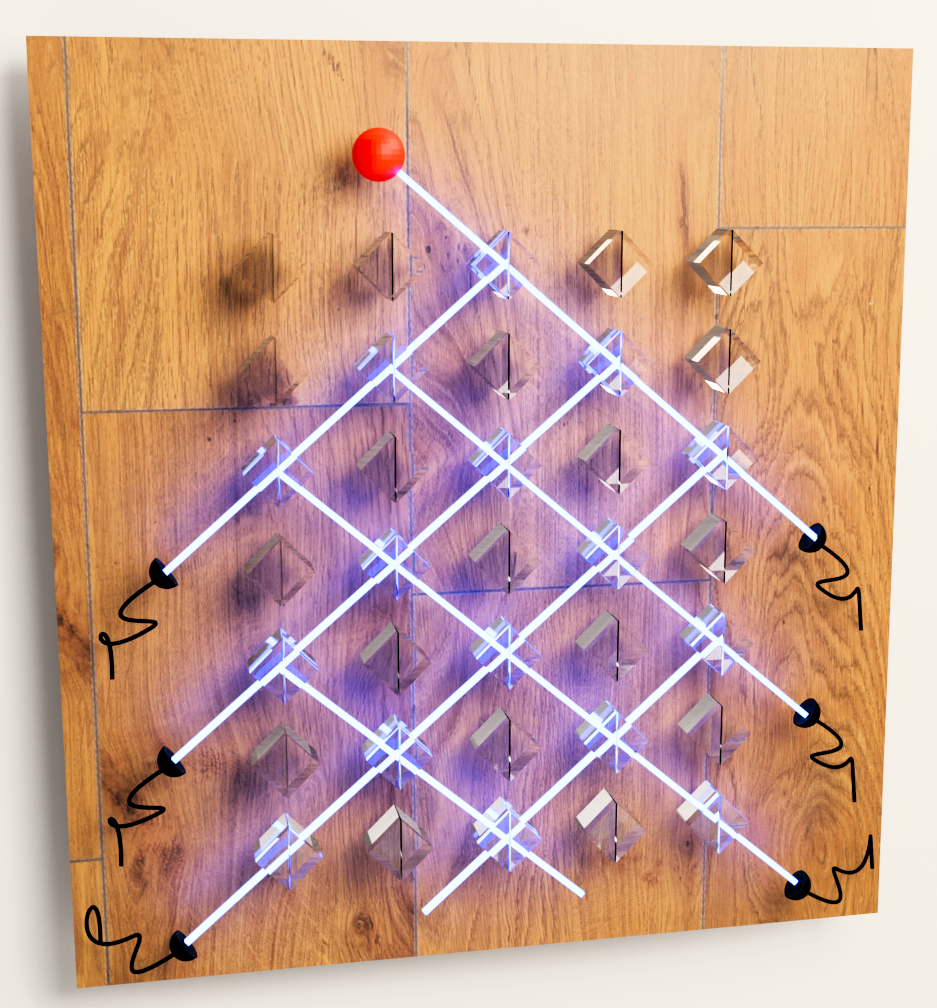}
\caption{ Quantum counterpart of the modified Galton board experiment scheme. Photons are shot from the top onto beamsplitters. At the ends of the board we have detectors. They measure at what time the photon went to the target position. Repeating this experiment with many particle will give us the average time of the photon being measured at target positions so the MHT.} 
\label{quantumgalton}
\end{figure}

In our one dimensional discrete time quantum walk to calculate MHT we use the setup proposed and numerically studied in \cite{Chelminiak2025}, with two measurement devices at the ends of the chain. Thus we extend the evolution by incorporating measurement with projection operator
\begin{equation}
    P = (\ket{x=x_{t_1}}\bra{x=x_{t_1}}+\ket{x=x_{t_2}}\bra{x=x_{t_2}})\otimes I_2.
\end{equation}
The evolution, assuming the particle has not been yet measured is given by
\begin{equation}\label{ew2}
    \ket{\Psi(t+1)} = W U\ket{\Psi(t)},
\end{equation}
with 
\begin{equation}
    W = I-P.
\end{equation}
Importantly while frequent measurements in a continuous-time limit could lead to the quantum Zeno effect, in our discrete-time scenario, the particle is able to propagate between measurements, ensuring a finite MHT. To gain better intuition about what MHT in quantum walks means we propose simple modification the the well known quantum counterpart of the Galton board. Similarly to classical walks the simplest experimental realization uses the Galton board, but instead of particle one uses photons and instead of wooden pins one puts beamsplitters. Additionally like in previous scenario to calculate MHT in this setup we propose modification in the form of detectors at both ends of the board which act the same as shelves on Fig. \ref{galton} and enable one to one correspondance with Eq. \ref{ew2}. The scheme of the experiment we show on Fig. \ref{quantumgalton}.

\section{Results}
Following the Krovi and Brun method of calculating MHT in discrete time quantum walk, we use their formula (see Appendix A, for derivation)
\begin{equation}\label{mht_krovi}
    \text{MHT} =\text{Tr} ( B(I-M)^{-2}\rho_0),
\end{equation}
with superoperators
\begin{equation}\label{b}
    B \rho = P U \rho \ U^\dagger P^\dagger,
\end{equation}
and
\begin{equation}\label{m}
    M \rho = W U \rho \ U^\dagger W^\dagger,
\end{equation}
and initial state
\begin{equation}
    \rho_0 = \ket{x=x_0,+1}\bra{x=x_0,+1}.
\end{equation}
Note that this can be solved if $(I-M)$ is not singular, which fortunately in our case $(I-M)$ is not singular. Solving Eq. (\ref{mht_krovi}) using symbolic programming for given $x_{t_1}$ and $x_{t_2}$ (starting from the middle) surprisingly yields the same result as in classical walks.
\begin{equation}\label{quantummht}
    \text{MHT} (x_0,x_{t_1},x_{t_2}) = |(x_{t_1}-x_0)(x_{t_2}-x_0)|,
\end{equation}
Thus we obtain analytical proof which contradicts common belief that in the context of this simple task quantum and classical walks achieve equal times.

Note that this was hinted by numerical experiments in \cite{Chelminiak2025}. They also showed that even though the MHT is the same in quantum and classical walks one still can show quantum advantage. However, one has to consider models with stochastic resetting which means that each step with probability $p$ the particle will be resetted to initial state. Let us assume that the initial state is symmetric relative to target positions. Then in the classical walk scenario the reset simply cannot help decrease the MHT, because it always moves the particle away from the destinations. Thus with the increase of reset probability $p$, the MHT of classical walk increases, diverging to infinity as $p$ goes to one. However, Che\l{}miniak et al. \cite{Chelminiak2025} demonstrated through numerical experiments that a stochastic reset can decrease the MHT for quantum walks. The authors supported this unexpected result with a heuristic argument based on prior work \cite{BACH2004562}, which showed that measurements at the edges of a chain act by reflecting a portion of the wave function even when the measurement outcome is null. It was further established that the reflective and absorbing components of the wave function can be distinguished by their quasi-momentum distribution. Consequently, after a measurement (reflection), the absorbing part of the wave function has already been absorbed, leaving the remaining wave function more reflective and less likely to be measured in subsequent steps. Che\l{}miniak et al. \cite{Chelminiak2025} correctly identified that a stochastic reset in a quantum scenario does not merely restore the particle to its initial position, it also redistributes the quasi-momenta in accordance with the uncertainty principle. This redistribution allows the particle to regain the more absorbing quasi-momenta, thereby facilitating a reduction in MHT. The observed dependence of MHT on the reset probability $p$, reflects a competition between two opposing effects: the reset returns the particle to the origin, which increases the MHT, but it simultaneously redistributes quasi-momenta, which decreases the MHT. Thus, depending on the distance to the target positions, the MHT decreases as p increases until an optimal reset probability is reached. Beyond this point, the MHT increases and eventually diverges to infinity at $p=1$, as the particle is reset so frequently that it can no longer move. We aim to provide analytical expressions for these dynamics by adapting the MHT derivation procedure developed by Krovi and Brun \cite{PhysRevA.74.042334} to accommodate quantum walks with stochastic reset.

\begin{figure}[t]
\includegraphics[width=7cm]{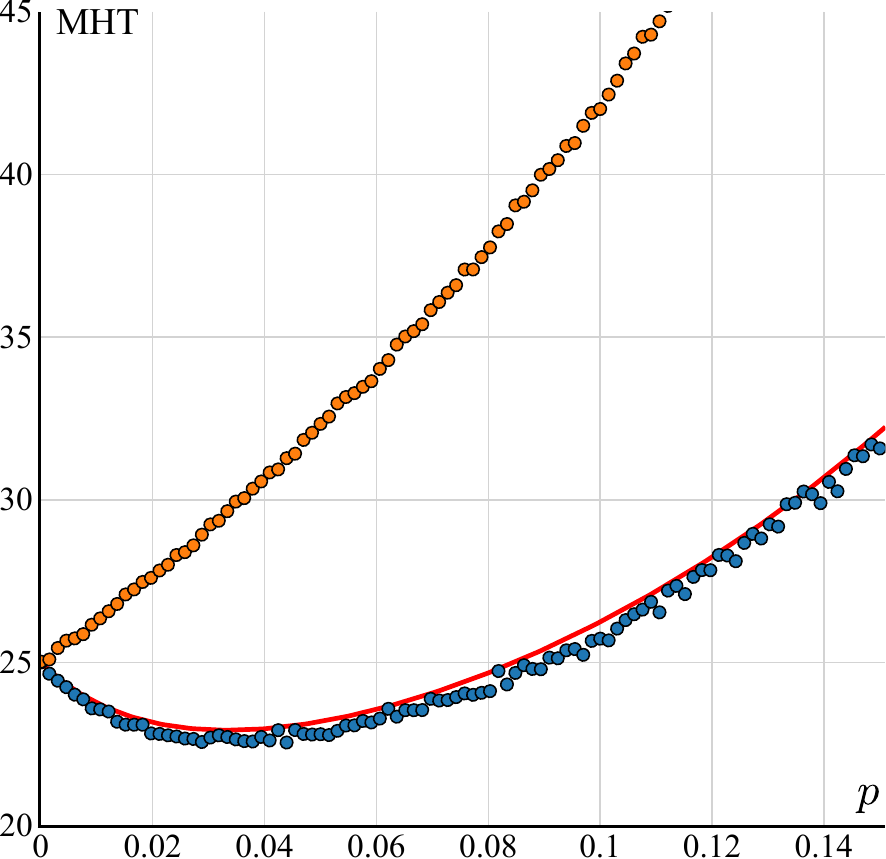}
\caption{ The MHT over reset probability $p$ for classical (orange) and quantum (blue) walk obtained in the numerical experiment with the result obtained by using symbolic programming to solve Eq. \ref{mht_krovi} (red line) (the analytical expression is shown in Appendix B). The walks started in $x=0$ and target positions were $x_{t_1}=-x_{t_2}=5$.}
\label{comp}
\end{figure}

Note that their procedure can be generalized to complete positive trace preserving maps (CPTP) instead of unitary evolution $U$. Thus we write the quantum walk with stochastic reset in the Krauss operator formalism. The evolution without measurements is given by 
\begin{equation}\label{Kraus}
    \rho(t+1) = \sum_i K_i \rho(t) K_i^\dagger,
\end{equation}
with $\sum_i K_i^\dagger K_i = I$. Let us take $K_0 = \sqrt{1-p} \ U$ which governs the usual quantum walk evolution, and 
\begin{equation}
    K_{j \in \langle1,2L\rangle} = \sqrt{p}\ket{\Psi(t=0)}\bra{j},
\end{equation}
that governs the reset (the analog of $2L$ dimensional extreme amplitude damping channel, which projects everything onto the initial state). Putting this Krauss evolution instead of $U$ in Eqs. (\ref{b},\ref{m}) we can obtain analytical result for the MHT of quantum walk with stochastic reset (see Fig. \ref{comp}). This proves that the reset can decrease the mean hitting time. Further, thanks to this approach we are able to obtain the optimal reset parameter which is challenging for numerical simulations (because of numerical errors in obtaining the MHT for quantum walks). 

Note that in this work we focus on the symmetric localized initial states. However, the proposed analytical approach is general and can be used to study scenarios with different initial states. As an example Che\l{}miniak et al. numerically studied the asymmetric state for which the MHT without reset is lower than for symmetric case (in accordance with Eq. \ref{quantummht}) but also the decrease of MHT with reset is lower. Another example would be the initial Gaussian distribution for which the finite size effects coming from the effective size of the chain come into play. The study of different initial states will be presented in later articles.

The proposed by us method of comparing the quantum and random walk needs one more detail. The quantum advantage gained from the reset procedure should decrease with noise. Specifically the quantum walk with bit flip error (with error probability 1/2) at internal degree of freedom should give the same MHT as the classical walk. 

\begin{figure}[t]
\includegraphics[width=7cm]{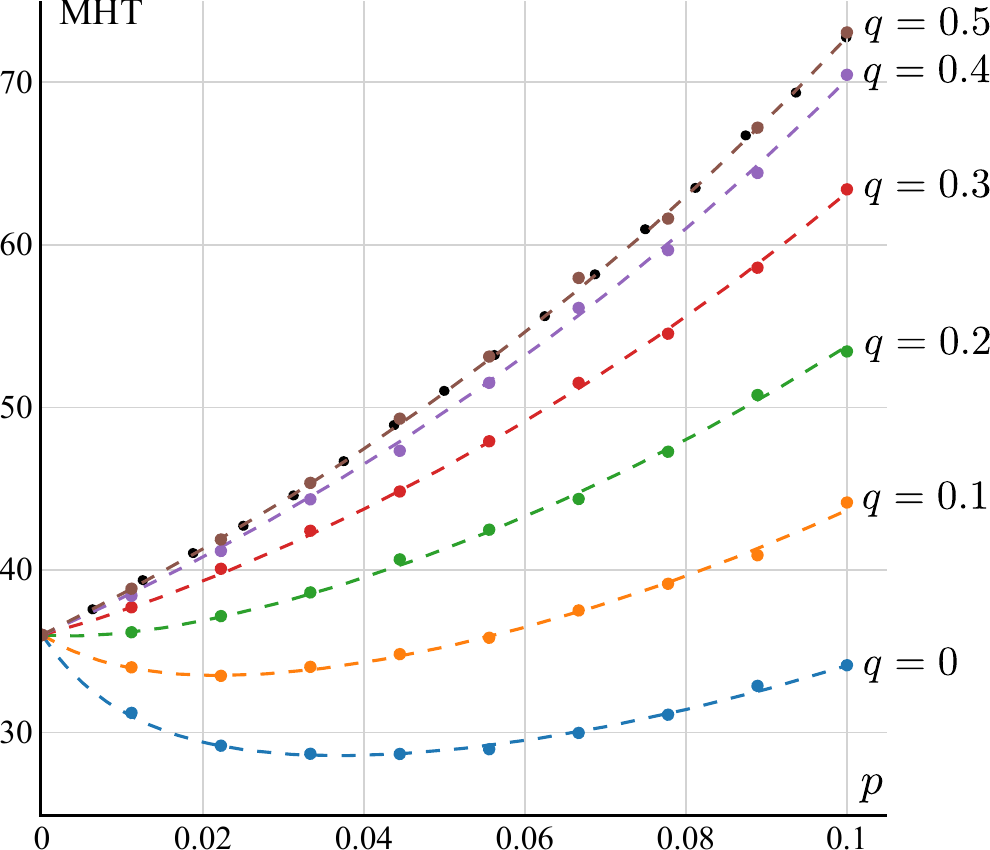}
\caption{ The MHT over reset probability $p$ for noisy quantum walk. The implemented noise was the bit flip error on the coin degree of freedom with probability $q$. The walks started in $x=0$ and target positions were $x_{t_1}=-x_{t_2}=6$. Additionally with black dots we put the results for classical random walk obtained from numerical experiment.}
\label{noise}
\end{figure}

The noise in quantum walk can be easily adapted through another Krauss evolution this time with

\begin{equation}
    \rho(t+1) = \sum_j \bar{K}_j \sum_i (K_i \rho(t) K_i^\dagger)\bar{K}_j,
\end{equation}
with 
\begin{equation}
    \bar{K}_0 = \sqrt{1-q} \ I,~~~~ \bar{K}_1 = \sqrt{q}\ I_{x}\otimes\sigma_x.
\end{equation}
For such noisy quantum walk we indeed obtain the intuitive result, the decrease of quantum advantage with the amplitude of noise (see Fig. \ref{noise}). 
For our noise analysis, we selected bit-flip errors on the coin degree of freedom as they provide a natural comparison point. At maximum strength ($q=1/2$), this noise effectively randomizes the quantum walk to match classical behavior, validating our MHT metric.

However, different noise models present distinct challenges. Consider amplitude damping on the internal degree of freedom: this creates a directional bias, causing the particle to preferentially move in one direction and artificially reducing the MHT. While such biased walks achieve low MHT values, this reflects the induced directionality rather than genuine quantum effects. These cases require careful interpretation, as the comparison between unbiased classical walks and biased quantum walks becomes less meaningful.

Despite these complications, the reset mechanism reveals an important insight. Even in biased quantum walks, any MHT reduction from stochastic reset must arise from quasi-momentum redistribution, a fundamentally quantum effect rooted in the uncertainty principle. Classical walks, whether biased or unbiased, cannot access this mechanism. Therefore, observing MHT decrease under reset remains a signature of quantum behavior, regardless of the specific noise model.

This observation has practical implications for quantum computing implementations. On quantum devices, where the walker's position is encoded across multiple qubits, even single-qubit errors can produce complex, non-local effects on the quantum walk. Standard noise models may not capture these complexities. Nevertheless, our reset-based metric provides a robust test: if stochastic reset reduces MHT in an experimental implementation, this demonstrates that quantum coherence effects remain active despite the noise. This offers a practical benchmark for assessing the "quantumness" of quantum walk implementations on real devices, complementing traditional fidelity measures.
\section{Discussion}
In this work, we have examined the conventional use of mean squared displacement for comparing quantum and classical random walks alongside an alternative metric. While MSD has been widely used to demonstrate quantum advantage and remains valuable for many purposes, we suggest that additional metrics can provide complementary insights: MSD is most meaningful for Gaussian distributions, yet quantum walk distributions are highly non-Gaussian.

We have shown that mean hitting time provides an alternative comparison that may be particularly relevant for certain applications. While the MSD gives the general information about the dynamics it can lead to misinterpretation where the MHT gives specific information about results in assigned task. Our results show that for the standard setup with two detectors, quantum and classical walks yield identical MHT when starting from symmetric initial conditions, suggesting neither model has an inherent advantage in the simplest scenario.

However, introducing stochastic resetting changes this picture dramatically. We have shown analytically that for quantum walks, the reset mechanism can decrease MHT, while classical walks see no benefit from resetting. This quantum advantage arises from quasi-momentum redistribution: different momentum components have different measurement probabilities, and reset allows the system to redistribute these components favorably.

We have further shown that this quantum advantage degrades appropriately with noise. For bit-flip errors approaching $q=1/2$, the quantum walk becomes randomized and MHT converges to the classical value, validating our metric. This leads us to propose that MHT decrease under stochastic reset serves as a signature of quantum behavior, particularly valuable for assessing quantum walk implementations on noisy quantum devices.

Our results suggest that quantum advantage in walk-based algorithms can manifest differently depending on the chosen metric, highlighting the importance of considering multiple measures, and that stochastic reset mechanisms may offer advantages even in moderately noisy environments.

The new analytical approach presented here not only proves the existence of a quantum advantage but is also crucial for harnessing it. This framework enables further systematic study and the search for optimal reset parameters. As hinted in \cite{Chelminiak2025}, the MHT depends significantly on the initial state and other walk parameters. Our method provides the tools necessary for such thorough examinations which we leave for further works.
\section{Data availability statement}
The data that support the finding of this article are openly available at \cite{data}

\clearpage
\onecolumngrid
\section{Appendix A}
In this section we show the derivation made by Krovi and Brun in \cite{PhysRevA.74.042334}. In discrete time quantum walk with measurement devices at the edges we define the mean hitting time as
\begin{equation}
    \text{MHT} = \sum_{t=0}^\infty tp(t),
\end{equation}
where $p(t)$ is probability of particle being ("absorbed") measured at either of the edges. Let us initialize the walk from some state $\rho_0$ then if till time $t$ the particle has not been measured its state is given by
\begin{equation}
    \rho(t) = (WU)^t \rho_0 (U^\dagger W)^t, 
\end{equation}
with $W = I-P$ where $P$ is the projective operator onto the measurement subspace. Thus the probability of measuring (a previously not measured particle) at time $t$ is given by
\begin{equation}
    p(t) = \text{Tr}\bigg(PU \ \rho(t-1)\ U^\dagger P\bigg) = \text{Tr}\bigg(P U (WU)^{t-1}\ \rho_0\ (U^\dagger W)^{t-1}U^\dagger P\bigg)
\end{equation}
or using superoperators
\begin{equation}
    p(t) = \text{Tr}\bigg(B\ M^{(t-1)} \rho_0\bigg).
\end{equation}
Substituting it into the MHT formula we get
\begin{equation}
    \text{MHT} = \text{Tr}\bigg(B\ \big(\sum_t tM^{(t-1)}) \ \rho_0\bigg),
\end{equation}
which (if $I-M$ is not singular) we can rewrite using Krovi and Brun \cite{PhysRevA.74.042334} trick as
\begin{equation}
    \text{MHT} = \text{Tr}\bigg(B\ (I-M)^{-2} \ \rho_0\bigg).
\end{equation}
\section{Appendix B}
Here we show the analytical expression for MHT$(p)$ for $x_{t_1} = -x_{t_2}=5$ which was obtained using symbolic programming to solve 
\begin{equation}
    \text{MHT}(p) =\text{Tr} ( B(p)(I-M(p))^{-2}\rho_0),
\end{equation}
with 
\begin{equation}
    B(p) \rho = \sum_i P K_i (p)\rho K_i^\dagger(p) P^\dagger,
\end{equation}
and
\begin{equation}
    M(p) \rho = \sum_i W K_i (p)\rho K_i^\dagger(p) W^\dagger,
\end{equation}
and
\begin{equation}
    \rho_0 = \ket{x=0,+1}\bra{x=0,+1},
\end{equation}
with $\sum_i K_i^\dagger K_i = I$ and $K_0 = \sqrt{1-p} \ U$, and
\begin{equation}
    K_{j \in \langle1,2L\rangle} = \sqrt{p}\ket{\Psi(t=0)}\bra{j}.
\end{equation}
 Finally we obtain result given by
\[
\text{MHT}(p) = \frac{P(p)}{Q(p)} =
\frac{\displaystyle \sum_{i=0}^{31} a_i p^i}
     {\displaystyle \sum_{i=0}^{29} b_i p^i},
\]
where the coefficients \(a_i\) and \(b_i\) are:

\begin{table*}[h!]
\centering
\small
\setlength{\tabcolsep}{6pt}
\renewcommand{\arraystretch}{1.1}
\begin{tabular}{c|rrrrrrr}
$i$   & 0 & 1 & 2 & 3 & 4 & 5 & 6 \\ \hline
$a_i$ & 25 & 680 & 3182 & $-16076$ & 41900 & 188272 & $-2365266$ \\
$b_i$ & 1 & 35 & 122 & $-1918$ & 5444 & 19864 & $-277774$ \\ \hline\hline
$i$   & 7 & 8 & 9 & 10 & 11 & 12 & 13 \\ \hline
$a_i$ & 13199288 & $-52780841$ & 168369748 & $-448139860$ & 1019082020 & $-2007144828$ & 3450802448 \\
$b_i$ & 1586138 & $-6286809$ & 19419993 & $-49059564$ & 103784468 & $-186111868$ & 284770388 \\ \hline\hline
$i$   & 14 & 15 & 16 & 17 & 18 & 19 & 20 \\ \hline
$a_i$ & $-5199757900$ & 6877143700 & $-7981402521$ & 8115959704 & $-7212800234$ & 5583320608 & $-3748159536$ \\
$b_i$ & $-373154260$ & 419592924 & $-405209569$ & 336023109 & $-238989182$ & 145441050 & $-75450536$ \\ \hline\hline
$i$   & 21 & 22 & 23 & 24 & 25 & 26 & 27 \\ \hline
$a_i$ & 2170309920 & $-1076624642$ & 453683956 & $-160649327$ & 47130156 & $-11239984$ & 2122056 \\
$b_i$ & 33180052 & $-12269534$ & 3771338 & $-947615$ & 189911 & $-29232$ & 3248 \\ \hline\hline
$i$   & 28 & 29 & 30 & 31 &  &  &  \\ \hline
$a_i$ & $-304968$ & 31328 & $-2048$ & 64 &  &  &  \\
$b_i$ & $-232$ & 8 &  &  &  &  &  \\
\end{tabular}
\caption{Coefficients \(a_i\) and \(b_i\) for the numerator and denominator of \(\text{MHT}(p)\).}
\end{table*}

\end{document}